\documentclass[pra,twocolumn,showpacs]{revtex4}

\usepackage{graphicx}
\usepackage{latexsym}
\usepackage{amsmath,amssymb}
\usepackage{verbatim,times,bbm}

\newtheorem{definition}{Definition}

\newtheorem{criterion}{Criterion}
\newtheorem{remark}{Remark}

\newcommand{\proofend}{\hfill\fbox\\\medskip }
\newcommand{\proof}[2]{{\noindent\bf Proof of #1} #2 $\proofend$}

\begin{document}

\title{The transitional behavior of quantum Gaussian memory channels}
\author{C. Lupo$^1$ and S. Mancini$^{1,2}$}
\affiliation{$^1$School of Science and Technology, University of
Camerino, I-62032 Camerino, Italy, EU\\
$^2$INFN-Sezione di Perugia, I-06123 Perugia, Italy, EU}

\begin{abstract}
We address the question of optimality of entangled input states in
quantum Gaussian memory channels. For a class of such channels,
which can be traced back to the memoryless setting, we state a
criterion which relates the optimality of entangled inputs to the
symmetry properties of the channels' action. Several examples of
channel models belonging to this class are discussed.
\end{abstract}

\pacs{03.67.Hk, 03.67.Mn}

\maketitle

\section{Introduction}

In communication theory, it is generally believed that memory
effects improve the information transfer capabilities of a
communication line \cite{Gall}. Memory effects can be introduced
both by the presence of correlations in the noise affecting
different channel uses (inputs), and by interference among theme.

In the quantum communication scenario the problem of determining the
optimal ensemble of input states, depending on the memory, naturally
arises. An optimal ensemble is the most robust under the action of
the noisy channel, leading to the highest transmission rates. In
particular, for the classical information transmission through
quantum memory channels, the possibility of discriminating between
the optimality of separable or entangled input states would be
useful.

Then, by the \emph{transitional behavior} it is intended the
possibility to single out two ``phases" in the channel capacity. One
for which the optimal input states are entangled among different
channel uses, and the other for which the optimal input states are
separable. To model the memory effects it is customary to introduce
a {\it memory parameter}, which quantifies the amount of
correlations in the quantum channel. The transition between the two
``phases" may happen at a finite value of the memory parameter,
implying that separable input states are optimal in the presence of
small correlations, or at zero value of the memory parameter,
implying that separable states are optimal only in the memoryless
limit. Several models showing such an effect have been proposed for
discrete quantum memory channels \cite{discretem} as well as for
continuous ones \cite{continuousm}. However, the majority of these
works restrict their analysis to few channel uses (thus not relying
on capacity arguments) and above all do not provide general
criterion to characterize the transitional behavior. Here we present
necessary conditions to have the transitional behavior for a class
of quantum Gaussian memory channels. Actually we show that such
behavior is intimately related to the symmetry properties of the
channels' action. This is possible thanks to the introduction of a
technique which allows us to {\it unravel} the memory effects.

The article is organized as follows. In Sec.\ \ref{G_tools} we
review the basic tools for working out quantum Gaussian memory
channels, and to the problem of evaluating the channel capacity. In
Sec.\ \ref{Holevo_Sec} we consider the problem of computing the
Holevo function for a Gaussian memory channel, and introduce a class
of Gaussian memory channels for which this problem can be traced
back to the memoryless setting. In Sec.\ \ref{symmetry_Sec} we
enunciate a criterion for the transitional behavior and we relate
the optimality of entangled inputs to the symmetry properties of the
channels' action. In Sec.\ \ref{examples} illuminating examples are
presented. Finally, Sec.\ \ref{conclusion} is for concluding
remarks.

\section{Gaussian states and Gaussian channels}\label{G_tools}

Gaussian quantum states and Gaussian quantum channels are defined in
the context of continuous variable quantum systems. Here we consider
the case of a continuous variable quantum system consisting of $n$
identical quantum harmonic oscillators (for a complete presentation
of the subject see for instance \cite{CV}). The $n$ quantum harmonic
oscillators are associated with a set of $2n$ canonical operators
$\hat q_1, \hat p_1$, $\hat q_2, \hat p_2$, $\dots$ $\hat q_n, \hat
p_n$, satisfying canonical commutation relations $[ \hat q_h, \hat
p_k ] = i \delta_{hk}$ (here and in what follows we assume
$\hbar=1$).

The state of the $n$ modes can be described, using the formalism of
density operator, as a certain density $\hat\rho_{n}$ defined in the
$n$-mode Fock space. However, we find it more convenient to work in
the Wigner function representation. Let us recall that the Wigner
function associated to a density $\hat\rho_{n}$ is defined as
\begin{equation}\label{Wigner}
W_{n}(\mathbf{q},\mathbf{p}) = \int d^n\mathbf{y} \langle \mathbf{q}
- \mathbf{y} | \hat\rho_{n} | \mathbf{q} + \mathbf{y} \rangle e^{2 i
\pi \mathbf{y}\cdot\mathbf{p}},
\end{equation}
where we have introduced the numerical vectors $\mathbf{q} := (q_1,
\dots q_n) $, $\mathbf{p} := (p_1, \dots p_n) $, $\mathbf{y} :=
(y_1, \dots y_n) \in\mathbb{R}^n$, and we have denoted
\begin{equation}
| \mathbf{q} \pm \mathbf{y} \rangle := \otimes_{k=1}^n | q_k \pm y_k
\rangle
\end{equation}
the joint (generalized) eigenstates of the `position' operators $\{
\hat q_k \}_{k=1,\dots n}$, i.e.\ $\hat q_k | \mathbf{q} \pm
\mathbf{y} \rangle = (q_k \pm y_k)| \mathbf{q} \pm \mathbf{y}
\rangle$, for all $k=1,\dots n$.

By definition, Gaussian states are those described by a Gaussian
Wigner function. In the $n$-mode scenario, the Wigner function of a
Gaussian state is a multivariate Gaussian function:
\begin{equation}
W_{n}(\mathbf{x}) =
\frac{\exp{\left[-\frac{1}{2}(\mathbf{x}-\mathbf{m})^\mathsf{T}
V^{-1}(\mathbf{x}-\mathbf{m})\right]}}{(2\pi)^n\sqrt{\det(V)}} ,
\end{equation}
where we have introduced the numerical vector
$\mathbf{x}:=(q_1,p_1,q_2,p_2,\dots q_n, p_n)^\mathsf{T} \in
\mathbb{R}^{2n}$. The Wigner function of an $n$-mode Gaussian state
is hence completely described by the vector of the $2n$ first
moments
\begin{equation}
\mathbf{m} = \langle \mathbf{x} \rangle = \int \mathbf{x}
W_{n}(\mathbf{x}) d^n\mathbf{x} \, ,
\end{equation}
and by the $2n \times 2n$ covariance matrix (CM)
\begin{equation}
V = \langle
(\mathbf{x}-\mathbf{m})(\mathbf{x}-\mathbf{m})^\mathsf{T} \rangle =
\int (\mathbf{x}-\mathbf{m})(\mathbf{x}-\mathbf{m})^\mathsf{T}
W_{n}(\mathbf{x}) d^n\mathbf{x} \, .
\end{equation}

Finally, let us notice that certain conditions have to be imposed on
the CM, to ensure that the Wigner function describes a {\it bona
fide} quantum state. Indeed, the Heisenberg principle imposes the
condition \cite{Mukunda}
\begin{equation}\label{unc}
V - i \Omega \ge 0,
\end{equation}
where
\begin{eqnarray}\label{symplecticform}
\Omega = \bigoplus_{k=1}^n \left(\begin{array}{cc} 0 & 1 \\ -1 & 0
\end{array}\right)
\end{eqnarray}
is the $2n \times 2n$ matrix representing the $n$-mode symplectic
form.

We will also consider an equivalent representation defined by a
different ordering of the canonical variables, expressed by the
numerical vector $ \mathbf{\tilde x} = (q_1, q_2, \dots q_n, p_1,
p_2, \dots p_n)^\mathsf{T}$, with the corresponding vector of first
moments $\mathbf{\tilde m}$ and the CM
\begin{equation}\label{alter}
\tilde V = \langle (\mathbf{\tilde x}-\mathbf{\tilde
m})(\mathbf{\tilde x}-\mathbf{\tilde m})^\mathsf{T} \rangle \, .
\end{equation}
In this representation, the symplectic form is represented by the
matrix
\begin{eqnarray}
\tilde\Omega = \left( \begin{array}{cc} \mathbb{O} & \mathbb{I}_n \\
-\mathbb{I}_n & \mathbb{O} \end{array} \right) \, ,
\end{eqnarray}
where $\mathbb{O}$ denotes the null matrix, and $\mathbb{I}_n$ the
identity matrix of size $n$.

A Gaussian quantum channel acting on $n$ bosonic modes (in short, an
$n$-mode Gaussian channel) is by definition a channel mapping
Gaussian states into Gaussian states. As a consequence, its action
on Gaussian states is completely characterized by the rule of
transformations of the vector of first moments and of the CM. One
can show (see, e.g., \cite{bgc}) that a Gaussian channel transforms
the pair $(\mathbf{m},V)$ (vector of first moment, CM) as follows:
\begin{equation}
(\mathbf{m},V) \mapsto ( X \mathbf{m} + \mathbf{d} , X V
X^\mathsf{T} + Y ).
\end{equation}
Where $\mathbf{d} \in \mathbb{R}^{2n}$ is a {\it displacement}
vector, and $X$, $Y$ are two $2n \times 2n$ matrices. In order to
represent a {\it bona fide} quantum channel, these matrices have to
obey the inequalities
\begin{equation}\label{cond_qc}
Y + i X \Omega X^\mathsf{T} - i \Omega \ge 0, \qquad Y \ge 0,
\end{equation}
In conclusion, a Gaussian channel is characterized by the triad
$(\mathbf{d},X,Y)$, satisfying Eq.\ (\ref{cond_qc}).

Given a pair of Gaussian channels: $\Phi$ with associated triad
$(\mathbf{d},X,Y)$, and $\Phi'$ with triad $(\mathbf{d}',X',Y')$,
the composition of the two channels $\Phi'\circ\Phi$ is associated
to the triad $(X'\mathbf{d}+\mathbf{d}',X'X,X'Y{X'}^\mathsf{T}+Y')$.

In the family of Gaussian channels, a special sub-family is those of
unitary Gaussian channels. Unitary Gaussian channels are
characterized by the conditions $Y=0$, and
\begin{equation}
X \Omega X^\mathsf{T} = \Omega.
\end{equation}
The last equation characterizes linear symplectic transformations,
where the matrix $X$ is symplectic. In conclusion, a Gaussian
unitary transformation is characterized by the triad
$(\mathbf{d},X,Y)$, where $\mathbf{d}$ is generic, $X$ is
symplectic, and $Y=0$. In the following we consider the subgroup of
Gaussian unitary transformations preserving the total number of
excitations, such maps are represented by matrices $X$ which are
both symplectic and orthogonal. Using the representation
(\ref{alter}), it is possible to show (see, e.g., \cite{CV} and the
references therein) that those matrices are of the form
\begin{eqnarray}\label{both}
\tilde X = \left(\begin{array}{cc}
\mathbb{A} & \mathbb{B} \\
-\mathbb{B} & \mathbb{A}
\end{array}\right) \, ,
\end{eqnarray}
where $\mathbb{A}$, $\mathbb{B}$ are real matrices of size $n$ such
that the matrix $\mathbb{A}+i\mathbb{B}$ is unitary.

\subsection{Gaussian memory channels, normal forms}

For any integer $n$, $n$ uses of a Gaussian channel transform $n$
input bosonic modes into $n$ output bosonic modes. The action of a
Gaussian quantum channel is hence described by a sequence of
Gaussian channels $\Phi_{n}$, acting on $n$ bosonic modes, which is
in turn associated to the sequence of triads
$(\mathbf{d}_{n},X_{n},Y_{n})$.

A very special case is that of the memoryless channels, for which
$\Phi_{n} = \Phi_{1}^{\otimes n}$ is the direct product of $n$
identical one-mode Gaussian channels. Each of these identical
one-mode Gaussian channels is characterized by a triad
$(\mathbf{d}_{1},X_{1},Y_{1})$. Hence, a memoryless channel is
characterized by a sequence $(\mathbf{d}_{n},X_{n},Y_{n})=
(\bigoplus_{k=1}^n \mathbf{d}_{1}, \bigoplus_{k=1}^n X_{1},
\bigoplus_{k=1}^n Y_{1})$. Notice that
$\mathbf{d}_1=(d_q,d_p)^\mathsf{T}$, and we have denoted
$\bigoplus_{k=1}^n \mathbf{d}_{1}:=$ $(d_q,d_p,$ $d_q,d_p,\dots$
$d_q,d_p)^\mathsf{T}$ $\in \mathbb{R}^{2n}$. Also notice that $X_1$,
$Y_1$ are $2 \times 2$ matrices, and the direct sums
$\bigoplus_{k=1}^n X_{1}$, $\bigoplus_{k=1}^n Y_{1}$ are $2n \times
2n$ block-diagonal matrices.

Let us now consider the case of a quantum channel with memory. We
call a {\it channel with memory}, or simply a {\it memory channel},
any channel which is {\it not} memoryless. Making no assumption on
additional structures that might be present (e.g.\ causality,
invariance under time translations), we can only say that the
associated sequence of $n$-mode Gaussian channels satisfies
$\Phi_{n} \neq \Phi_{1}^{\otimes n}$, and the associated sequence of
triads is $(\mathbf{d}_{n},X_{n},Y_{n}) \neq (\bigoplus_{k=1}^n
\mathbf{d}_{1}, \bigoplus_{k=1}^n X_{1}, \bigoplus_{k=1}^n Y_{1})$.

The problem of finding {\it normal forms} for $n$-mode Gaussian
channels was considered in \cite{Wolf,Caruso}, the case of one-mode
channel was considered in \cite{Holevo}. Normal forms are
equivalence classes of $n$-mode Gaussian channels, up to (Gaussian)
unitary equivalence. Hence, given a pair of $n$-mode Gaussian
channels $\Phi_n$, $\Phi'_n$, they are equivalent if there are
Gaussian unitary transformations $\mathcal{E}_n$, $\mathcal{D}_n$
such that $\Phi'_n = \mathcal{D}_n \circ \Phi_n \circ \mathcal{E}_n$
($\circ$ denotes the composition of channels). As the chosen
notation suggests, we want to look at $\mathcal{E}_n$,
$\mathcal{D}_n$ respectively as unitary {\it encoding} and {\it
decoding} transformations, $\mathcal{E}_n$ anticipating and
$\mathcal{D}_n$ following the action of the quantum channel. We will
hence write $(\mathbf{d}_n,X_n,Y_n)\simeq(\mathbf{d}'_n,X'_n,Y'_n)$
if the two Gaussian quantum channels are equivalent in the sense
declared above.

The first thing to be noticed is that
$(\mathbf{d}_n,X_n,Y_n)\simeq(0,X_n,Y_n)$. This is a well known
result, a consequence of the fact that the {\it displacement} vector
can always be eliminated by applying a proper $n$-mode displacement
operator (which is a Gaussian unitary transformation) at the
encoding, or decoding, stage. For this reason, in what follows we
only consider $n$-mode Gaussian channels of the form $(0,X_n,Y_n)$.
Let us consider a quantum channel $\Phi_n$ with the triad
$(0,X_n,Y_n)$, an encoding $\mathcal{E}_n$ with triad $(0,E_n,0)$,
and decoding $\mathcal{D}_n$ with triad $(0,D_n,0)$. The application
of the encoding and decoding unitaries leads to the {\it dressed}
channel $\Phi'_n$ associated to the triad $(0,D_n X_n E_n, D_n Y_n
D_n^\mathsf{T})$.

As was shown in \cite{Wolf}, an $n$-mode Gaussian channel is always
unitary equivalent to a $n$-mode channel in a {\it normal form},
i.e.\ $(0,X_n,Y_n)\simeq(0,X'_n,Y'_n)$, where
\begin{equation}\label{normal}
X'_n = \left[ \bigoplus_{h=1}^p X_{2}^{(h)} \right] \bigoplus \left[
\bigoplus_{k=2p+1}^n X_{1}^{(k)} \right]
\end{equation}
is the direct sum of $p$ two-mode ($4 \times 4$) matrices
$X_{2}^{(h)}$, and $n-2p$ one-mode ($2 \times 2$) matrices
$X_{1}^{(k)}$. In other words, by applying suitable encoding and
decoding unitaries, the matrix $X_n$ is reduced to the direct sum of
two-mode and one-mode terms. However, it is important to notice that
the matrix $Y'_n$ cannot be in general jointly reduced to the same
form.

\subsection{Classical capacity of Gaussian channels}

A quantum channel can be used to transmit classical information by
encoding a classical stochastic continuous variable $Z$, distributed
according to a probability density distribution $p_Z$, into a set of
quantum states $\hat\rho_Z$.

In the case of a memoryless quantum channel $\Phi_n =
\Phi_1^{\otimes n}$, the maximum rate at which classical information
can be reliably sent through the quantum channel is given by the
regularized limit \cite{HSW}
\begin{equation}\label{Holevo}
C = \lim_{n\to\infty} \frac{1}{n} \chi\left( \Phi_1^{\otimes n}
\right) \, ,
\end{equation}
where the Holevo function $\chi$ evaluated on $n$ channel uses is
\begin{align}
\chi\left( \Phi_1^{\otimes n} \right) = \max_{\{\hat\rho_Z, p_Z\}} &
\left\{ S\left[ \Phi_1^{\otimes n}\left( \int dZ p_Z \hat\rho_Z
\right) \right] \right.\nonumber\\
& \qquad \left. - \int dZ p_Z S\left[ \Phi_1^{\otimes
n}\left(\hat\rho_Z\right) \right] \right\} \, ,
\end{align}
where $S$ denotes the von Neumann entropy. If the von Neumann
entropy is expressed in qubits, then the classical capacity of the
quantum channel in measured in bits per channel use. The computation
of the memoryless channel capacity is based on the optimization over
all input ensembles, including those made of states which are
entangled among different channel uses. On the other hand, if the
input states are restricted to ensembles of separable states, one
obtains the so-called one-shot capacity
\begin{equation}
C_1 = \chi\left( \Phi_1 \right) \, .
\end{equation}
Clearly the one-shot capacity is a lower bound on the memoryless
channel capacity. If the two quantities coincide, the Holevo
function is said to be additive. Additivity of the Holevo function
dramatically simplifies the problem of evaluating the memoryless
channel capacity. Even though the Holevo function has been shown to
be additive for several relevant channels, e.g.\ the lossy channel
in the framework of Gaussian channels \cite{broadband}, this
property does not hold in general \cite{Hastings}. Non-additivity of
the Holevo function implies that the optimal ensembles of input
states, i.e.\ the most robust to noise, are entangled among
different channel uses: a phenomenon which has no counterpart in the
classical theory of information.

Moving to the case of quantum channels with memory, characterized by
the inequality $\Phi_n \neq \Phi_1^{\otimes n}$, one could be
tempted to generalize the formula in Eq.\ (\ref{Holevo}) and write
\begin{equation}\label{Holevo_m}
C \simeq \lim_{n\to\infty} \frac{1}{n} \chi\left( \Phi_n \right) \,
,
\end{equation}
with
\begin{align}
\chi\left( \Phi_n \right) = \max_{\{\hat\rho_Z, p_Z\}} & \left\{
S\left[ \Phi_n \left( \int dZ p_Z \hat\rho_Z \right) \right]
\right.\nonumber\\
& \qquad \left.- \int dZ p_Z S\left[ \Phi_n \left(\hat\rho_Z\right)
\right] \right\} \, .
\end{align}
Indeed, it is possible to show \cite{Mancini} that the right hand
site of Eq.\ (\ref{Holevo_m}) is in general only an upper bound for
the classical capacity of the memory channel. On the other hand, it
has been proven that this quantity coincides with the memory channel
capacity for the class of so-called {\it forgetful} channels
\cite{KW2}. Those channels have the property that correlations among
channels uses decay exponentially. Moreover, they exhibit a causal
structure and are invariant under time translation.

The problem of computing the classical capacity is extremely hard.
Indeed, if one cannot rely on the additivity property, one has to
evaluate the regularized limit of the Holevo information, whose
complexity increases exponentially in $n$. Moreover, for the case of
bosonic channels, the relevant Hilbert space is infinite dimensional
even for a single channel use, i.e.\ $n=1$. Clearly, an infinite
dimensional Hilbert space could carry an infinite amount of
classical bits. Hence, to avoid unphysical results, one is led to
introduce a physically motivated constraint, and to exploit the
concept of constrained capacity. A typical choice in the framework
of bosonic Gaussian channels is to impose a constraint on the
maximal input energy per channel use, hence one introduces the
constrained Holevo function
\begin{align}
\chi^N\left( \Phi_n \right) = \max_{\{\hat\rho_Z, p_Z\}} & \left\{
S\left[ \Phi_n \left( \int dZ p_Z \hat\rho_Z \right) \right]
\right.\nonumber\\
- & \int dZ p_Z S\left[ \Phi_n \left(\hat\rho_Z\right) \right]\nonumber\\
| \, \sum_{k=1}^n & \left. \mathrm{tr}\left( \frac{\hat q_k^2 + \hat
p_k^2}{2n} \int dZ p_Z \hat\rho_Z \right) \le N + \frac{1}{2}
\right\} \, , \label{chi_constr}
\end{align}
where we have assumed unit frequency for the bosonic oscillators,
and $N$ represents the maximum number of excitations per mode on
average. One can conjecture that the maximum is reached in
correspondence with a Gaussian ensembles. Indeed, the optimality of
the Gaussian ensemble has been proven for the lossy channel
\cite{broadband} and conjectured for other families of bosonic
Gaussian channels \cite{conjecture}.

Here, we estimate the Holevo function when restricted to Gaussian
encoding \cite{HW}. We consider Gaussian encoding defined as
follows. For $n$ uses of the quantum channel, we fix a reference
$n$-mode Gaussian state, with zero mean, described by the Wigner
function
\begin{equation}
W(\mathbf{x}) = \frac{\exp{\left[ -\frac{1}{2}\mathbf{x}^\mathsf{T}
V_\mathrm{in}^{-1}\mathbf{x}\right]}}{(2\pi)^n\sqrt{\det{(V_\mathrm{in})}}}
\, .
\end{equation}
A classical variable $\mathbf{m} \in \mathbb{R}^{2n}$ is hence
encoded by applying a displacement operation on the reference state,
thus obtaining
\begin{equation}
W_\mathbf{m}(\mathbf{x}) = \frac{\exp{\left[
-\frac{1}{2}(\mathbf{x}-\mathbf{m})^\mathsf{T}
V_\mathrm{in}^{-1}(\mathbf{x}-\mathbf{m})\right]}}{(2\pi)^n\sqrt{\det{(V_\mathrm{in})}}}
\, .
\end{equation}
We assume the stochastic variable $\mathbf{m}$ to be itself
distributed according to the Gaussian probability density
distribution with zero mean:
\begin{equation}
p_\mathbf{m} = \frac{\exp{\left[ -\frac{1}{2}\mathbf{m}^\mathsf{T}
V_\mathrm{c}^{-1} \mathbf{m}
\right]}}{(2\pi)^n\sqrt{\det{(V_\mathrm{c})}}} \, .
\end{equation}
By linearity of the relation (\ref{Wigner}), the corresponding
ensemble state
\begin{equation}
\int d^{2n}\mathbf{m} p_\mathbf{Z} \hat\rho_\mathbf{m}
\end{equation}
is itself described by a Gaussian Wigner function, i.e.\
\begin{equation}
\int d^{2n}\mathbf{m} p_\mathbf{m} W_\mathbf{m}(\mathbf{x}) =
\frac{\exp{\left[ -\frac{1}{2}\mathbf{x}^\mathsf{T}
(V_\mathbf{in}+V_\mathrm{c})^{-1} \mathbf{x}
\right]}}{(2\pi)^n\sqrt{\det{(V_\mathbf{in}+V_\mathrm{c})}}} \, .
\end{equation}
The restriction to Gaussian states, which are mapped into Gaussian
states by Gaussian channels, dramatically simplifies the problem,
since the complexity of Gaussian states is polynomial in the number
of modes $n$. Moreover, the von Neumann entropy can be calculated in
terms of the symplectic eigenvalues of the CM.

The symplectic eigenvalues of an $n$-mode CM $V$ are defined as
follows. Notice that the matrix $V\Omega$, where $\Omega$ is the
symplectic form introduced in Eq.\ (\ref{symplecticform}), has $2n$
purely imaginary eigenvalues $\{ \pm i \nu_k \}_{k=1,\dots n}$,
where the $n$ real numbers $\{ \nu_k \}_{k=1,\dots n}$ are the
symplectic eigenvalues of the CM $V$. From the uncertainty
relations, expressed by Eq.\ (\ref{unc}), it follows that the
symplectic eigenvalues satisfy the inequalities $\nu_k \ge 1/2$,
which are saturated by pure Gaussian states \cite{CV}. The von
Neumann entropy of an $n$-mode Gaussian state $\hat\rho$,
characterized by a CM $V$, is given by the formula:
\begin{equation}
S[\hat\rho] = \sum_{k=1}^n g\left(\nu_k-\frac{1}{2}\right)\, ,
\end{equation}
where we have introduced the function $g$ defined by
\begin{equation}
g(x) = (x+1)\log_2{(x+1)} - x \log_2{(x)} \, .
\end{equation}
The von Neumann entropy of a Gaussian state $\hat\rho$ is determined
by its CM $V$, it is hence convenient to define a function $\Sigma$
of the CM such that
\begin{equation}
\Sigma[V] = S(\hat\rho) \, .
\end{equation}

Finally, we notice that the input energy constraint in Eq.\
(\ref{chi_constr}) can be written in terms of the CM as follows:
\begin{equation}
\frac{\mathrm{tr}(V_\mathrm{in} + V_\mathrm{c})}{2n} \le N +
\frac{1}{2} \, .
\end{equation}

\section{The Holevo function for Gaussian ensembles} \label{Holevo_Sec}

For $n$ uses of the quantum channel, the constrained Holevo
function, when restricted over Gaussian ensembles, reads
\begin{align}
\chi_G^N(\Phi_n) = \max_{V_\mathrm{in},V_\mathrm{c}} \frac{1}{n} &
\left\{ \Sigma\left[ X_n (V_\mathrm{in}+V_\mathrm{c}) X_n^\mathsf{T}
+ Y_n \right] \right.\nonumber\\
& - \Sigma\left[ X_n V_\mathrm{in} X_n^\mathsf{T} + Y_n
\right]\nonumber\\
 | \, & \left. \frac{\mathrm{tr}(V_\mathrm{in} + V_\mathrm{c})}{2n} \le N +
\frac{1}{2} \right\} \, , \label{chi_G_n}
\end{align}
where the optimization is over the CMs $V_\mathrm{in}$,
$V_\mathrm{c}$ satisfying the energy constrains.

In the case of a memoryless Gaussian channel, by restricting on
Gaussian input states which are separable among different channel
uses, we get to the one-shot capacity:
\begin{align}
\chi_G^N(\Phi_1) = \max_{V_\mathrm{in},V_\mathrm{c}} & \left\{
\Sigma\left[ X_1 (V_\mathrm{in}+V_\mathrm{c}) X_1^\mathsf{T} + Y_1
\right] \right.\nonumber\\
& - \Sigma\left[ X_1 V_\mathrm{in} X_1^\mathsf{T} + Y_1 \right] \nonumber\\
| \, & \left.\frac{\mathrm{tr}(V_\mathrm{in} + V_\mathrm{c})}{2} \le
N + \frac{1}{2} \right\} \, . \label{chi_G_1}
\end{align}

In general, for a Gaussian memory channel, characterized by the
sequence of triads $(\mathbf{d}_n, X_n, Y_n )$, the optimization of
the $n$-use Holevo function in Eq.\ (\ref{chi_G_n}) cannot be
reduced to the one-use case in Eq.\ (\ref{chi_G_1}). That is a
consequence of the the normal form for an $n$-mode Gaussian channel
in Eq.\ (\ref{normal}), which in general is not in the form the
product of $n$ independent one-mode Gaussian channels. However, we
can still identify a class of Gaussian memory channels such that,
for any $n$, there are Gaussian unitary encoding and decoding
transformations $\mathcal{E}_n = (0,E_n,0)$, $\mathcal{D}_n =
(0,D_n,0)$, such that
\begin{equation}\label{dressing}
\mathcal{D}_n \circ \Phi_n \circ \mathcal{E}_n = \bigotimes_{k=1}^n
\Phi_1^{(k)}\, ,
\end{equation}
where $\{ \Phi_1^{(k)} \}_{k=1,\dots n}$ is a collection of $n$ (not
necessarily identical) one-mode Gaussian channels. In other words,
the memory channels belonging to that class factorize in terms of
the collective input and output variables defined by the encoding
and decoding transformations. Hence we introduce the following

\begin{definition}[memory unraveling]
A bosonic Gaussian memory channel, characterized by a sequence
$(0,X_n,Y_n)$, can be unraveled if there is a sequence of encoding
Gaussian unitaries $(0,E_n,0)$, and a sequence of decoding Gaussian
unitaries $(0,D_n,0)$, such that, for any $n$,
\begin{eqnarray}
D_{n}X_{n}E_{n} & = & \bigoplus_{k=1}^n X_1^{(k),n}, \label{EDX}\\
D_{n}Y_{n}D_{n}^\mathsf{T} & = & \bigoplus_{k=1}^n Y_1^{(k),n}.
\label{EDY}
\end{eqnarray}
Moreover, we require that the encoding unitary preserves the form of
the energy constraint.
\end{definition}

Since the Holevo function is invariant under unitary
transformations, the transformation (\ref{dressing}) preserves the
Holevo function, i.e.\
\begin{equation}
\chi_G^N\left[\Phi_n\right] =
\chi_G^N\left[\bigotimes_{k=1}^n\Phi_1^{(k)}\right] \, .
\end{equation}
Moreover, since the encoding Gaussian unitary preserves the form of
the energy constraint, we have
\begin{equation}
\mathrm{tr}\left[ E_n(V_\mathrm{in}+V_\mathrm{c})E_n^\mathsf{T}
\right] = \mathrm{tr}\left[ V_\mathrm{in}+V_\mathrm{c} \right] \, ,
\end{equation}
which is satisfied if $E_n^\mathsf{T} E_n = 1$, i.e.\ if the matrix
$E_n$ is orthogonal. Symplectic matrices that are also orthogonal
constitute a subgroup whose elements are characterized by the form
in Eq.\ (\ref{both}).

For a memory channel that can be unraveled, it is natural to
restrict to Gaussian input ensembles such that
\begin{equation}
E_n V_\mathrm{in} E_n^\mathsf{T} = \bigoplus_{k=1, \dots n} V_{1,
\mathrm{in}}^{(k)} \, ,
\end{equation}
and
\begin{equation}
E_n V_\mathrm{c} E_n^\mathsf{T} = \bigoplus_{k=1, \dots n} V_{1,
\mathrm{c}}^{(k)} \, .
\end{equation}
Using this ansatz, the calculation of the Holevo function for $n$
uses of the memory channel reduces to the one-mode case:
\begin{align}
\chi_G^N\left[\Phi_n\right] = & \frac{1}{n} \max_{\{ N_k | \sum_k
N_k/n = N \}} \sum_{k=1}^n \max_{\{
V_{1,\mathrm{in}}^{(k)},V_{1,\mathrm{c}}^{(k)}\}} \nonumber\\
& \left\{ \Sigma\left[ X_1^{(k),n}
(V_{1,\mathrm{in}}^{(k)}+V_{1,\mathrm{c}}^{(k)})
{X_1^{(k),n}}^\mathsf{T} + Y_1^{(k),n} \right] \right. \nonumber \\
& - \Sigma\left[ X_1^{(k),n} V_{1,\mathrm{in}}^{(k)}
{X_1^{(k),n}}^\mathsf{T} + Y_1^{(k),n} \right] \nonumber\\
& \, | \, \left.\frac{\mathrm{tr}(V_{1,\mathrm{in}}^{(k)} +
V_{1,\mathrm{c}}^{(k)})}{2} \le N_k + \frac{1}{2} \right\} \, ,
\label{Holevo_G}
\end{align}
where we have rewritten the input energy constraint in two steps
\begin{eqnarray}
\left\{\begin{array}{ccc}
\mathrm{tr}\left(V_1^{(k),n}+V_{1,\mathrm{c}}^{(k)}\right)/2 & \le & N_k + 1/2 \, , \\
\sum_{k=1}^n N_k/n & = & N \, ,
\end{array}\right.
\end{eqnarray}
and the maximization is over both the CMs $V_{1,\mathrm{in}}^{(k)}$,
$V_{1,\mathrm{c}}^{(k)}$, and over the positive integers $N_k$ under
the constraint $\sum_k N_k/n = N$.

In conclusion, for quantum memory channels that can be unraveled,
the calculation of the Holevo function has been reduced to the case
of independent, but not identical, one-mode channels, each
characterized by the matrices $X_1^{(k),n}$, $Y_1^{(k),n}$ and a
maximal number $N_k$ of excitations per mode. The only ingredient
that mixes the one-mode channels is the constraint on the total
number of excitations.

\subsection{Examples}\label{Exs1}

The properties defining the class of Gaussian memory channels that
can be unraveled are rather peculiar, however several relevant
models of Gaussian channels belong to this class. In this section we
review some examples of Gaussian memory channels that can (or
cannot) be unraveled.

{\it Lossy Bosonic memory channel.} We refer to the general model of
lossy bosonic Gaussian channel with memory that has been introduced
in \cite{GiovManc}. Upon $n$ uses of the channel, $n$ input modes
are mixed with a corresponding set of $n$ environmental modes at a
beam splitter with transmissivity $\eta$. In the Heisenberg picture
the canonical field operators transform as
\begin{eqnarray}
\hat q_k & \rightarrow & \sqrt{\eta} \, \hat q_k + \sqrt{1-\eta} \, \hat Q_k \\
\hat p_k & \rightarrow & \sqrt{\eta} \, \hat p_k + \sqrt{1-\eta} \,
\hat P_k \, ,
\end{eqnarray}
where $\{ \hat Q_k, \hat P_k \}_{k=1,\dots n}$ are the field
operators of the environmental modes. The environmental modes are in
a correlated Gaussian state, which is characterized by a CM
$V_\mathrm{env}$ with nonvanishing off-diagonal terms coupling
different modes. The memory channel is associated with the matrices
$X_n = \sqrt{\eta} \mathbb{I}_{2n}$, $Y_n = (1-\eta)
V_\mathrm{env}$. Memory effects appear in the channel if the
$n$-mode environment is in a non-factorized state. A factorized
state is characterized by a block-diagonal $n$-mode CM, i.e.\
$V_\mathrm{env}=\bigoplus_{k=1}^n v_k$. Remarkable examples of
non-factorized states are the multimode entangled states, which
belong to the family of multimode squeezed states. To address the
problem of memory unraveling, we notice that the CM $V_\mathrm{env}$
can be diagonalized by a $2n \times 2n$ orthogonal matrix $O$, i.e.\
$O V_\mathrm{env} O^\mathsf{T} = \bigoplus_{k=1}^n v_k$. Hence one
could identify $D_n:=O$, $E_n:=O^\mathsf{T}$. Notice that the
orthogonal matrix preserves the trace, hence it preserves the energy
constraint. However, a Gaussian unitary is represented with a
symplectic matrix, hence the given orthogonal matrix represents a
physical transformation only if it is also symplectic. In
conclusion, a lossy bosonic memory channel can be unraveled if the
CM of the environment can be diagonalized by a linear transformation
which is both symplectic and orthogonal. A characterization of this
class of environment CMs is presented in \cite{Lupo}: pure Gaussian
states and squeezed thermal states belong to this class.

{\it Additive noise channel.} For this class of channels, the field
operators transforms according to the Heisenberg picture map:
\begin{eqnarray}
\hat q_k & \rightarrow & \hat q_k + t_k \, , \\
\hat p_k & \rightarrow & \hat p_k + u_k \, ,
\end{eqnarray}
where $t_k$, $u_k$ are classical stochastic variables. The channel
is Gaussian if the noise variables are Gaussian distributed, with a
CM $V_\mathrm{cl}$. The subscript '$\mathrm{cl}$' refers to the fact
that $V_\mathrm{cl}$ is a classical CM, i.e.\ it is only subject to
the conditions of being symmetric and positive semi-definite.
Differently from the quantum CM, it needs not obey the Heisenberg
uncertainty relations as expressed by Eq.\ (\ref{unc}). Memory
effects arise when the $V_\mathrm{cl}$ has nonvanishing off-diagonal
terms coupling different channel uses. The matrices associated to
$n$ uses of the channel are $X_n = \mathbb{I}_{2n}$, and $Y_n =
V_\mathrm{cl}$. The memory channel can hence be unraveled if there
is a $2n \times 2n$ matrix $S$, being both symplectic and
orthogonal, such that $S_n V_\mathrm{cl} S_n^\mathsf{T} =
\bigoplus_{k=1}^n v_k$. The encoding and decoding Gaussian unitaries
which unravel the memory channel are hence chosen as $D_n := S_n$,
$E_n := S_n^\mathsf{T}$. Notice that since $S_n$ is orthogonal,
$S_n^\mathsf{T}=S_n^{-1}$. Examples of additive noise channels with
memory that can be unraveled were studied in \cite{Markov,Markov_2}.
The conditions on the CM $V_\mathrm{cl}$ can be easily obtained as
it is done for the lossy channel in \cite{Lupo}, with the only
difference that $V_\mathrm{cl}$ needs not obey the uncertainty
relations.

{\it Inter-symbol interference channels.} In this family of quantum
channels memory effects come from the fact that the signals at
different channel inputs do interfere at the channel output, while
in the previous examples they are caused by noise correlations. For
such a case, one can write the Heisenberg picture transformations
acting on the field operators as
\begin{eqnarray}
\hat q_k & \rightarrow & \sum_h \left\{\mathbb{M}_{kh}^{(qq)} \hat
q_h + \mathbb{M}_{kh}^{(qp)} \hat
p_h + \mathbb{N}_{kh}^{(qQ)} \hat Q_k + \mathbb{N}_{kh}^{(qP)} \hat P_h \right\} \, , \nonumber\\
\hat p_k & \rightarrow & \sum_h \left\{ \mathbb{M}_{kh}^{(pq)} \hat
q_h + \mathbb{M}_{kh}^{(pp)} \hat p_h + \mathbb{N}_{kh}^{(pQ)} \hat
Q_k + \mathbb{N}_{kh}^{(pP)} \hat P_h \right\} \, , \nonumber
\end{eqnarray}
where the operators $\{ \hat Q_k, \hat P_k \}$ correspond to
environmental modes, and the matrices $\mathbb{M}^{(qq)}$,
$\mathbb{M}^{(qp)}$, $\dots$, $\mathbb{N}^{(pQ)}$,
$\mathbb{N}^{(pP)}$ satisfy proper conditions \cite{CV}. To describe
this channel, we work in the representation (\ref{alter}). In this
representation, assuming that the environmental modes are in a
Gaussian state characterized by the CM $\tilde V_\mathrm{env}$, the
matrices associated with the memory channel are
\begin{eqnarray}
\tilde X_n = \left(\begin{array}{cc}
\mathbb{M}^{(qq)} & \mathbb{M}^{(qp)} \\
\mathbb{M}^{(pq)} & \mathbb{M}^{(pp)}
\end{array}\right) \, ,
\end{eqnarray}
and
\begin{eqnarray}
\tilde Y_n = \left(\begin{array}{cc}
\mathbb{N}^{(qQ)} & \mathbb{N}^{(qP)} \\
\mathbb{N}^{(pQ)} & \mathbb{N}^{(pP)}
\end{array}\right)
\tilde V_\mathrm{env}
\left(\begin{array}{cc}
\mathbb{N}^{(qQ)} & \mathbb{N}^{(qP)} \\
\mathbb{N}^{(pQ)} & \mathbb{N}^{(pP)}
\end{array}\right)^\mathsf{T} \, .
\end{eqnarray}
An instance of this kind of model was considered in \cite{inter}, in
which $\mathbb{M}^{(qq)}=\mathbb{M}^{(pp)}$,
$\mathbb{M}^{(qp)}=\mathbb{M}^{(pq)}=\mathbb{O}$,
$\mathbb{N}^{(qQ)}=\mathbb{N}^{(pP)}$,
$\mathbb{N}^{(qP)}=\mathbb{N}^{(pQ)}=\mathbb{O}$, and the
environment is in the vacuum state, i.e.\ $\tilde V_\mathrm{env} =
\mathbb{I}_{2n}/2$. As is shown in \cite{inter}, such a channel can
be unraveled by performing the {\it singular value decomposition} of
the matrix $X_n$.

\section{Optimization under symmetries}\label{symmetry_Sec}

In the cases of both memoryless and memory quantum channels, one can
pose the question of finding the optimal input ensemble, i.e.\ the
most robust one under the action of the noisy channel. In the
memoryless setting, this is related to the issue of additivity of
the Holevo information: if the Holevo information is additive the
optimal input ensemble constitutes of states which are separable
among different channel uses; otherwise, for channels with a
non-additive Holevo information the optimal input ensemble is made
of entangled states. For memory channels it has been observed, in
the cases of both discrete \cite{discretem} and continuous
\cite{continuousm} variables, that entangled input states may be
optimal to maximize the Holevo information. In models for quantum
channels with memory, is customary to introduce a {\it memory
parameter}, used to {\it quantify} the memory in the channel, which
vanishes in the memoryless limit. It may happen that the optimal
input ensemble constitutes of entangled states when the memory
parameter is above a certain threshold. In this case one says that
the memory channel exhibits a {\it transitional behavior}. In the
case of discrete variables, several models exhibit a finite value of
the memory threshold \cite{discretem}, while for continuous-variable
models the threshold value may vanish \cite{continuousm,Markov_2},
i.e.\ entangled input states are optimal even for arbitrary small,
but not zero, values of the memory parameter.

Restricting to the case of bosonic Gaussian channels that can be
unraveled, here we introduce a criterion to decide whether the
Holevo information, restricted on Gaussian input ensembles and under
input energy constraint, is optimized by separable input states. We
formulate the following:

\begin{criterion}\label{cr1}
Given a Gaussian memory channel, represented by the sequence
$(0,X_n,Y_n)$, and provided that it can be unraveled, a necessary
condition for the optimality of entangled input states is the non
invariance under phase rotation.
\end{criterion}

\proof{Criterion \ref{cr1}}{For a Gaussian memory channel that can
be unraveled, it holds $(0,X_n,Y_n)\simeq(0,\bigoplus_{k=1,\dots
n}X_1^{(k),n}, \bigoplus_{k=1,\dots n}Y_1^{(k),n} )$. Let us assume,
by contradiction, that the one-mode channels $(0, X_1^{(k),n},
Y_1^{(k),n} )$ are invariant under phase rotation, i.e.\
\begin{eqnarray}
R_1(\theta) X_1^{(k),n} R_1(\theta)^\mathsf{T} & = & X_1^{(k),n}, \label{RX}\\
R_1(\theta) Y_1^{(k),n} R_1(\theta)^\mathsf{T} & = & Y_1^{(k),n},
\label{RY}
\end{eqnarray}
where the $2 \times 2$ matrix $R_1(\theta)$ represents a phase
rotation
\begin{eqnarray}
R_1(\theta) := \left(\begin{array}{cc}
\cos{\theta} & -\sin{\theta} \\
\sin{\theta} & \cos{\theta}
\end{array}\right).
\end{eqnarray}
Matrices that have this symmetry are scalar, i.e.\
\begin{eqnarray}
X_1^{(k),n} & = & \left(\begin{array}{cc}
x^{(k),n} & 0 \\
0 & x^{(k),n}
\end{array}\right) \, , \\
Y_1^{(k),n} & = & \left(\begin{array}{cc}
y^{(k),n} & 0 \\
0 & y^{(k),n}
\end{array}\right)\, .
\end{eqnarray}

It follows that the solution of the optimization problem
(\ref{Holevo_G}) is also invariant under phase rotation and is given
by the matrices:
\begin{eqnarray}\label{symmetric}
V_{1,\mathrm{in}}^{(k)} = \left(\begin{array}{cc}
1/2 & 0 \\
0 & 1/2
\end{array}\right)\, , \quad
V_{1,\mathrm{c}}^{(k)} = \left(\begin{array}{cc}
N_k & 0 \\
0 & N_k
\end{array}\right) \, ,
\end{eqnarray}
where the optimal values of the parameters $\{N_k\}_{k=1,\dots n}$
are obtained from the maximization problem
\begin{align}
\chi_n = \max_{\{N_k\}} \frac{1}{n} \sum_{k=1}^n & \left\{
g\left[(x^{(k),n})^2(N_k+1/2)+y^{(k),n}-1/2\right]
\right.\nonumber\\
&\left. - g\left[(x^{(k),n})^2(1/2)+y^{(k),n}-1/2\right] \right\} \,
, \label{chi_n}
\end{align}
where the maximum is under the constraint $\sum_{k=1}^n N_k/n = N$.

Let us notice that the matrix $V_{1,\mathrm{in}}^{(k)}$ in
(\ref{symmetric}) represents the CM of coherent states \cite{CV}.
Then, the matrix $\bigoplus_{k=1}^n V_1^{(k),n}$ represents the CM
of the optimal $n$-mode input state of the {\it dressed} memory
channel, which includes the encoding unitary transformation. The
actual optimal input state is obtained from it by undoing the
encoding transformation, i.e.\
\begin{equation}
V_\mathrm{in}^\mathrm{opt} = E_n \left[ \bigoplus_{k=1}^n
V_1^{(k),n} \right] E_n^\mathsf{T}.
\end{equation}
However, since the encoding matrix $E_n$ is orthogonal (and, of
course symplectic), it follows that
\begin{eqnarray}
V_\mathrm{in}^\mathrm{opt} = \bigoplus_{k=1}^n
\left(\begin{array}{cc}1/2 & 0
\\ 0 & 1/2\end{array}\right),
\end{eqnarray}
i.e.\ the optimal Gaussian inputs are coherent states, which are
separable among different channel uses.}

% % % % % % % % % % % % % % % % % % % % % % % % % % % % % % % % % % % % % % % % % % % %

\begin{remark}\label{rmrk1}
If the decoding symplectic matrix $D_n$ is also orthogonal, the
conditions (\ref{RX}), (\ref{RY}) can be equivalently formulated as
follows:
\begin{eqnarray}
R_n(\theta) X_n R_n(\theta)^\mathsf{T} & = & X_n, \\
R_n(\theta) Y_n R_n(\theta)^\mathsf{T} & = & Y_n,
\end{eqnarray}
where the matrix $R_n(\theta)=\bigoplus_{k=1}^n R_1(\theta)$
represents a global phase rotation on the $n$ modes.
\end{remark}

The equivalence can be readily proved by working in the
representation (\ref{alter}). It is a consequence of the fact that
the subgroup of symplectic and orthogonal matrices [having the form
as in Eq.\ (\ref{both})] commutes with phase rotations, which are
represented by matrices of the following form:
\begin{eqnarray}\label{phase_tilde}
\tilde R_n = \left( \begin{array}{cc}
\cos{\theta} \mathbb{I}_n & -\sin{\theta} \mathbb{I}_n \\
\sin{\theta} \mathbb{I}_n & \cos{\theta} \mathbb{I}_n
\end{array}\right) \, .
\end{eqnarray}

In other words, if both the encoding and decoding symplectic
matrices are orthogonal, the symmetry under phase rotation can be
checked directly on the matrices $X_n$, $Y_n$.

Furthermore, being the $D_n$, $E_n$ elements of Lie groups, they
reduce to identity when the group's parameters reduce to zero. Since
the latter would characterize the degree of memory, we may argue
that the \emph{transition} can only occur at zero value of the
memory parameters. This is in contrast to what happens in discrete
quantum memory channels.

As a consequence of the Criterion \ref{cr1}, entangled Gaussian
codewords may be necessary for optimizing the Holevo information
(\ref{Holevo_G}) only if the rotational invariance is broken. Notice
however that this is a necessary but not sufficient condition. Below
we are going to present examples for all possible cases.

\section{Examples}\label{examples}

Let us first consider the case of an additive noise channel. The
definition of the channel and its basic properties are briefly
recalled in Sec.\ \ref{Exs1}. Let us recall that for $n$ uses of the
channel we have $X_n = \mathbb{I}_{2n}$, and $Y_n = V_\mathrm{cl}$.
Since we are considering channels which can be unraveled, we assume
the existence of an orthogonal and symplectic matrix $S_n$, such
that $S_n V_\mathrm{cl} S_n^\mathsf{T}=\bigoplus_{k=1}^n v_k$, with
$D_n=S_n$, $E_n=S_n^\mathsf{T}$. In this case, the symmetry
condition is verified if the matrix $V_\mathrm{cl}$ is symmetric
under phase rotation (remark \ref{rmrk1}). Two models of Gaussian
memory channels with Markovian correlated noise were studied and
characterized in \cite{Markov,Markov_2}. Using the representation
(\ref{alter}), the noise CM in \cite{Markov} has the following form
\begin{eqnarray}
\tilde V_\mathrm{cl} = \left( \begin{array}{cc}
\mathbb{V} & \mathbb{O} \\
\mathbb{O} & \mathbb{V} \end{array}\right) \, ,
\end{eqnarray}
which is clearly symmetric under phase rotations
(\ref{phase_tilde}), hence the optimal Gaussian inputs are separable
for this model. On the contrary, the noise CM for the model studied
in \cite{Markov_2} has the form
\begin{eqnarray}
\tilde V_\mathrm{cl} = \left( \begin{array}{cc}
\mathbb{V} & \mathbb{O} \\
\mathbb{O} & \mathbb{V}' \end{array}\right) \, ,
\end{eqnarray}
with $\mathbb{V}\neq\mathbb{V}'$. Such a matrix is not symmetric
under phase rotations (\ref{phase_tilde}) and, as shown in
\cite{Markov_2}, the optimal input states, when restricted to
Gaussian states, are entangled.

The case of lossy bosonic memory channel \cite{GiovManc} is
analogous to the additive channel. Its basic properties are reviewed
in Sec.\ \ref{Exs1}. In this case $X_n = \sqrt{\eta}
\mathbb{I}_{2n}$, and $Y_n=(1-\eta)V_\mathrm{env}$. A model of
memory channel belonging to this family has been studied in
\cite{Lupo}. Using the representation (\ref{alter}), the
environmental CM in \cite{Lupo} has the form
\begin{eqnarray}
\tilde V_\mathrm{env} = \left(T+\frac{1}{2}\right)\left(
\begin{array}{cc}
e^{\mathbb{M}s} & \mathbb{O} \\
\mathbb{O} & e^{-\mathbb{M}s} \end{array}\right) \, ,
\end{eqnarray}
where $\mathbb{M}$ is a symmetric matrix of size $n$, and $T$, $s$
are two positive parameters. The parameter $s$ quantifies the amount
of memory in the channel: For $s=0$ the environmental state is an
uncorrelated thermal state, while it is entangled for $s \neq 0$.
For all $s \neq 0$, the matrix $Y_n$ is not symmetric under phase
rotation and, as shown in \cite{Lupo}, the optimal Gaussian input
states are entangled among different channel uses.

The general case of an inter-symbol interference channel is recalled
in Sec.\ \ref{Exs1}. An example of such a channel was studied and
characterized in \cite{inter}, where
\begin{eqnarray}\label{X_matrix}
\tilde X_n = \left(\begin{array}{cc}
\mathbb{M} & \mathbb{O} \\
\mathbb{O} & \mathbb{M}
\end{array}\right) \, ,
\end{eqnarray}
and
\begin{eqnarray}
\tilde Y_n = \frac{1}{2} \left(\begin{array}{cc}
\mathbb{N}\mathbb{N}^\mathsf{T} & \mathbb{O} \\
\mathbb{O} & \mathbb{N}\mathbb{N}^\mathsf{T}
\end{array}\right) \, .
\end{eqnarray}
The channel is invariant under phase rotation, thus, as shown in
\cite{inter}, the optimal input states are separable. To conclude,
we notice that there are models of Gaussian memory channel for which
the optimal Gaussian input states are separable, even though the
channel is not symmetric under phase rotation. A channel model
presenting this feature can be obtained by choosing the environment
to be in a state of the form
\begin{eqnarray}
\tilde V_\mathrm{env} = \left(\begin{array}{cc}
e^s \mathbb{I}_n & \mathbb{O} \\
\mathbb{O} & e^{-s} \mathbb{I}_n
\end{array}\right) \, ,
\end{eqnarray}
implying
\begin{eqnarray}
\tilde Y_n = \frac{1}{2} \left(\begin{array}{cc}
e^s \mathbb{N}\mathbb{N}^\mathsf{T} & \mathbb{O} \\
\mathbb{O} & e^{-s} \mathbb{N}\mathbb{N}^\mathsf{T}
\end{array}\right) \, ,
\end{eqnarray}
and choosing $\tilde X_n$ as in (\ref{X_matrix}). The channel is
hence not symmetric under phase rotation, however it is not
difficult to show that the optimal Gaussian input states are
separable.

\section{Conclusion}\label{conclusion}

In this article we have provided a unified framework for some recent
results about the performance of quantum Gaussian memory channels.
We have focused in particular on the entanglement of optimal input
states and we have related this issue to the symmetry properties of
the channel. More specifically we have shown that entangled Gaussian
codewords might be necessary for optimizing the Holevo information
only if the rotational invariance is broken by the channel's action.
Similar considerations were also done in
\cite{Markov,Markov_2,CerfPRA} for specific channel models.
Moreover, for a Gaussian memory channel that can be unraveled, we
may argue that the transition from the optimality of separable
states to the optimality of entangled states may only occur for
vanishing value of the memory parameter \cite{Lupo,Markov_2}. This
is in contrast to what happens in discrete quantum memory channels
\cite{discretem}. However, while there investigations have only
involved very few channel uses, here the analysis has been carried
out for arbitrary number of channel uses by resorting to a
mathematical machinery called memory unraveling. That allowed us to
trace the Gaussian memory channel back to a memoryless one. Several
examples have been discussed concerning memory unraveling as well as
transitional behavior.

We think that the presented results shed light on the mechanisms and
the structure of the correlations that lead to an enhancement of the
channel performance with entanglement, although a complete
characterization of memory channels transition features is still far
away.

\acknowledgments

This work has been supported by the European Commission under the
FET-Open grant agreement CORNER, number FP7-ICT-213681.


\begin{thebibliography}{99}

\bibitem{Gall}
R. G. Gallager, \emph{Information Theory and Reliable
Communication} (Wiley, New York, 1968).

\bibitem{discretem}
C. Macchiavello and G. M. Palma, Phys. Rev. A \textbf{65}, 050301(R)
(2002); E. Karpov, D. Daems and N.J. Cerf, Phys. Rev. A \textbf{74},
032320 (2006); D. Daems, Phys. Rev. A \textbf{76}, 012310 (2007); F.
Caruso, V. Giovannetti, C. Macchiavello and M.B. Ruskai, Phys. Rev.
A \textbf{77}, 052323 (2008).

\bibitem{continuousm}
G. Ruggeri, G. Soliani, V. Giovannetti and S. Mancini, Europhys.
Lett. \textbf{70}, 719 (2005); G. Ruggeri and S. Mancini, Quantum
Inf. Comput. \textbf{7}, 265 (2007); O. Pilyavets, V. Zborovskii and
S. Mancini, Physs Rev. A \textbf{77}, 052324 (2008).

\bibitem{CV} A. Ferraro, S. Olivares and M. G. A. Paris, \textit{Gaussian states in
quantum information} (Bibliopolis, Napoli, 2005); eprint
arXiv:0503237 [quant-ph]; S. L. Braunstein, P. van Loock, Rev. Mod.
Phys. {\bf 77}, 513 (2005).

\bibitem{Mukunda} R. Simon, N. Mukunda, and B. Dutta, Phys.
Rev. A {\bf 49}, 1567 (1994).

\bibitem{bgc} A. S. Holevo, R. F. Werner, Phys. Rev. A {\bf 63},
032312 (2001).

\bibitem{Wolf} M. M. Wolf, Phys. Rev. Lett. {\bf 100}, 070505
(2008).

\bibitem{Caruso} F. Caruso, J. Eisert, V. Giovannetti and A. S.
Holevo, New J. Phys. {\bf 10}, 083030 (2008).

\bibitem{Holevo} A. S. Holevo, arXiv:quant-ph/0607051.

\bibitem{HSW} A. S. Holevo, IEEE Trans. Inf. Theory {\bf 44}, 269 (1998); B. Schumacher, M. D. Westmoreland, Phys. Rev. A {\bf 56}, 131 (1997).

\bibitem{broadband}V. Giovannetti, S. Guha, S. Lloyd, L. Maccone, J. H. Shapiro, H. P. Yuen, Phys. Rev. Lett. {\bf 92}, 027902
(2004).

\bibitem{Hastings} M. B. Hastings, Nature Physics {\bf 5}, 255 (2009).

\bibitem{Mancini} G. Bowen, S. Mancini, Phys. Rev. A {\bf 69}, 012306 (2004).

\bibitem{KW2} D. Kretschmann, R. F. Werner, Phys. Rev. A {\bf 72},
062323 (2005).

\bibitem{conjecture} V. Giovannetti, S. Guha, S. Lloyd, L. Maccone, J. H. Shapiro, Phys. Rev. A {\bf 70},
032315 (2004).

\bibitem{HW} A. S. Holevo, R. F. Werner Phys. Rev. A {\bf 63}, 032312
(2001); A. S. Holevo, M. Sohma, O. Hirota, Phys. Rev. A {\bf 59},
1820 (1999).

\bibitem{GiovManc} V. Giovannetti and S. Mancini, Phys. Rev. A {\bf 71}, 062304 (2005).

\bibitem{Lupo} C. Lupo, O. V. Pilyavets, S. Mancini, New J. Phys. {\bf 11}, 063023 (2009).

\bibitem{Markov} C. Lupo, L. Memarzadeh, S. Mancini, Phys. Rev. A {\bf
80}, 042328 (2009).

\bibitem{Markov_2} J. Sch\"afer, D. Daems, E. Karpov, N. J. Cerf, Phys. Rev. A {\bf
80}, 062313 (2009).

\bibitem{inter} C. Lupo, V. Giovannetti, S. Mancini, Phys. Rev. Lett. {\bf 104}, 030501
(2010).

\bibitem{CerfPRA} N. J. Cerf, J. Clavareau, C. Macchiavello, J. Roland, Phys. Rev. A {\bf 72}, 042330 (2005).

\end{thebibliography}
\end{document}